\documentclass[aps,pra,showpacs,superscriptaddress,preprint]{revtex4}
\usepackage{amssymb}
\usepackage{amsmath}
\usepackage{graphicx}

\catcode`ð=\active
 \defð{\u{g}}
 \catcode`Ð=\active
 \defÐ{\u{G}}
 \catcode`Ý=\active
\defÝ{\. I}
 \catcode`ö=\active
\defö{\"{o}}
 \catcode`Ö=\active
 \defÖ{\"O}
 \catcode`ü=\active
 \defü{\"{u}}
 \catcode`Ü=\active
 \defÜ{\"{U}}
 \catcode`Þ=\active
\defÞ{\c{S}}
 \catcode`þ=\active
 \defþ{\c{s}}
 \catcode`ý=\active
 \defý{{\i}}
 \catcode`ç=\active
\defç{\d{c}}
 \catcode`Ç=\active
\defÇ{\d{C}}

\input{tcilatex}

\begin{document}

\title{Spin and pseudospin symmetry along with orbital dependency of the
Dirac-Hulth$\mathbf{{\acute{e}}}$n problem}
\author{Sameer M. Ikhdair}
\email[E-mail: ]{sikhdair@neu.edu.tr}
\affiliation{Physics Department, Near East University, Nicosia, North Cyprus, Turkey}
\author{C\"{u}neyt Berkdemir}
\email[E-mail: ]{berkdemir@erciyes.edu.tr}
\affiliation{Physics Department, Erciyes University, 38039 Kayseri, Turkey}
\author{Ramazan Sever}
\email[E-mail: ]{sever@metu.edu.tr}
\affiliation{Physics Department, Middle East Technical University, 06531 Ankara, Turkey}
\date{%
\today%
}

\begin{abstract}
The role of the Hulth$\mathbf{{\acute{e}}}$n potential on the spin and
pseudospin symmetry solutions is investigated systematically by solving the
Dirac equation with attractive scalar $S(\vec{r})$ and repulsive vector $V(%
\vec{r})$ potentials. The spin and pseudospin symmetry along with orbital
dependency (pseudospin-orbit and spin-orbit dependent couplings) of the
Dirac equation are included to the solution by introducing the Hulth$\mathbf{%
{\acute{e}}}$n-square approximation. This effective approach is based on
forming the spin and pseudo-centrifugal kinetic energy term from the square
of the Hulth$\mathbf{{\acute{e}}}$n potential. The analytical solutions of
the Dirac equation for the Hulth$\mathbf{{\acute{e}}}$n potential with the
spin-orbit and pseudospin-orbit-dependent couplings are obtained by using
the Nikiforov-Uvarov (NU) method. The energy eigenvalue equations and wave
functions for various degenerate states are presented for several
spin-orbital, pseudospin-orbital and radial quantum numbers under the
condition of the spin and pseudospin symmetry.

\end{abstract}

\maketitle

\newpage

\section{Introduction}

The spin and pseudospin symmetry [1,2] observed originally almost 40 years
ago as a mechanism to explain different aspects of the nuclear structure is
one of the most interesting phenomena in the relativistic quantum mechanics.
It plays a crucial role for a Dirac hamiltonian with realistic scalar $S(%
\vec{r})$ and vector $V(\vec{r})$ potentials, for nucleon spectrum in
nuclei, for the existence of identical bands in superdeformed nuclei, etc
[3]. The key feature of the pseudospin symmetry is based on the small energy
difference between single-nucleon doublets with quantum numbers $n_{r},\ell
,j=\ell +1/2$ and $n_{r}-1,\ell +2,j=\ell +3/2$, where $n_{r}$, $\ell $ and $%
j$ are the single nucleon radial, orbital and total angular quantum numbers,
respectively. These quantum numbers are relabelled as pseudospin doublets; $%
\ell +1=\tilde{\ell}$ is the "pseudo" orbital angular momentum, $\tilde{s}%
=1/2$ is the "pseudo" spin and $j=\tilde{\ell}\pm \tilde{s}$ is the total
"pseudo" angular momentum for the two states in the doublet [4]. For
example, "$n_{r}s_{1/2},(n_{r}-1)d_{3/2}$" is valid for $\tilde{\ell}=1$, "$%
n_{r}p_{3/2},(n_{r}-1)f_{5/2}$ is valid for $\tilde{\ell}=2$, etc. Another
key feature is the single-particle Hamiltonian of the oscillator shell
model. This means that the pseudospin concept in the nuclear theory is a
division of the single-particle total angular momentum into \textit{pseudo}
rather than \textit{normal} orbital and spin parts. The shell model implies
that nucleons move in a relativistic mean field produced by the interactions
between nucleons. The relativistic dynamics of nucleons moving in the
relativistic mean field are described by using the Dirac equation and not
the Schr\"{o}dinger equation.

The pseudospin symmetry concept is investigated by the framework of the
Dirac equation and occurs as a symmetry of the Dirac hamiltonian when an
attractive scalar $S(\vec{r})$ and a repulsive vector $V(\vec{r})$
potentials near equal to each other in magnitude, but opposite in sign,
i.e., $S(\vec{r})\sim -V(\vec{r})$. On the other hand, the sum of the vector
and scalar potentials in the Dirac equation is a constant, \textit{i.e.}, $V(%
\vec{r})+S(\vec{r})=constant$, for the solution of the pseudospin symmetry
in nuclei. This condition has been found by Ginocchio [5] and applied to the
case of the spherical harmonic oscillator [6]. Meng \textit{et al} [7]
showed that the pseudospin symmetry is exact under the condition of $d(V(%
\vec{r})+S(\vec{r}))/dr=0$. Lisboa \textit{et al.} studied the generalized
harmonic oscillator for spin-$1/2$ particles by setting either $\Sigma (\vec{%
r})=V(\vec{r})+S(\vec{r})=0$ or $\Delta (\vec{r})=V(\vec{r})-S(\vec{r})=0$
[8]. A necessary condition for occurrence of the pseudospin symmetry in
nuclei is to consider the case $\Sigma (\vec{r})=0$ [3,5,9,10]. For more
realistic nuclear systems, the quality of the pseudospin symmetry is
increased in the framework of the single-particle relativistic models and
hence the competition between the pseudo-centrifugal barrier and the
pseudospin-orbital potential is completed in the onset of pseudospin
symmetry [11]. The Dirac equation with the pseudospin symmetry is solved
numerically for nucleons which move independently in the relativistic mean
field with external scalar and vector potentials [12,13]. In addition to the
numerical solutions, some analytical solutions are also discussed for
solving the Dirac equation for some realistic potentials [14-17] with the
pseudospin symmetry. The analytical solutions show that under the condition
of pseudospin symmetry, the exact solution of the Dirac equation gives the
bound-state energy spectra and spinor wave functions [18-20].

The aim of this paper is to present an analytical bound state solutions of
the Dirac equation for the Hulth$\mathbf{{\acute{e}}}$n potential under the
conditions of the exact pseudospin symmetry and exact spin symmetry. To
obtain a general solution for all values of the pseudospin (spin) quantum
numbers, the pseudospin (spin) symmetry and orbital dependency,
pseudospin-orbit (spin-orbit) dependent coupling are included to the lower
component of the Dirac equation as an integer quantum number. This component
has the structure of the Schr\"{o}dinger-like equations with the
pseudo-centrifugal (spin-centrifugal) kinetic energy term and its solution
is analyzed by using some algebraic methods and effective approaches. One of
these effective approaches is applied to the pseudo-centrifugal
(spin-symmetry) kinetic energy term in the case of $\tilde{\ell}>0$ ($\ell
>0 $) and also an effective potential suggested in the form of the square of
the Hulth$\mathbf{{\acute{e}}}$n potential is taken into account instead of
the pseudo-centrifugal kinetic energy term. For small values of the radial
coordinate $r$, this effective potential gives a centrifugal energy term in
the first approximation. Therefore, the pseudo-centrifugal
(spin-centrifugal) kinetic energy term is accepted as an effective term in
this region. It is worthy to state that Jia \textit{et al} [21,22] have
proposed an improved new approximation scheme to deal with the centrifugal
kinetic energy term in the solution of the Schr\"{o}dinger-Hulth$\mathbf{{%
\acute{e}}}$n problem. Using this approximation scheme, Jia \textit{et al}
[23,24] have obtained approximate analytical solutions for the
Dirac-generalized P\"{o}schl-Teller and Klein-Gordon-P\"{o}schl-Teller
problems including the centrifugal kinetic energy term. Recently, Ikhdair
[25] has applied the approximation scheme to deal with the orbital
centrifugal term in the Schr\"{o}dinger-Manning-Rosen problem using the
Nikiforov-Uvarov method. Further, the approximation has also been applied to
the Schr\"{o}dinger--Hulth$\mathbf{{\acute{e}}}$n problem using the improved
quantization rule [26].

In the present work, the Dirac equation for the Hulth$\mathbf{{\acute{e}}}$n
potential is arranged under the condition of the exact pseudospin (spin)
symmetry and it's solution is obtained systematically by using the
Nikiforov-Uvarov (NU) method [27-31]. As an application of the Dirac-Hulth$%
\mathbf{{\acute{e}}}$n problem with the pseudospin (spin) symmetry, the
relativistic eigenvalue spectrum for various degenerate states is presented
for several pseudo-orbital (spin-orbital) and pseudospin (spin) quantum
numbers.

The structure of the paper is as follows. In Sec. 2, the basic ideas of the
Nikiforov-Uvarov (NU) method are outlined in short. In Sec. 3, the Dirac
equation is briefly introduced for the spin and pseudospin symmetry
solutions. In Sec. 4, the Hulth$\mathbf{{\acute{e}}}$n potential is
substituted into the lower component of the Dirac equation and the
pseudo-centrifugal (or spin-centrifugal) kinetic energy term is replaced by
the square of the Hulth$\mathbf{{\acute{e}}}$n potential to apply the Hulth$%
\mathbf{{\acute{e}}}$n square approximation. The main results obtained in
previous sections are connected by means of the main equation of the NU
method. Lastly, the general procedures of the solution method are followed
to obtain the energy eigenvalue equation and two-spinor wave functions.
Results and conclusions are performed in Sec. 5.

\section{Basic Ideas of the Nikiforov-Uvarov (NU) Method}

It is especially well known that the solutions of the Schr\"{o}dinger and
Schr\"{o}dinger-like equations including the centrifugal barrier and/or the
spin-orbit coupling terms have not been obtained straightforwardly for the
exponential-type potentials such as Morse, Hulth$\mathbf{{\acute{e}}}$n,
Woods-Saxon, etc [32]. Although the exact solution of the Schr\"{o}dinger
equation for the exponential-type potentials has been obtained for $\ell =0$%
, any $\ell $-state solutions have been given approximately by using some
analytical methods under a certain number of restrictions [33,34]. One of
the calculational tools utilized in these studies is the NU method. This
technique is based on solving the hypergeometric type second-order
differential equations by means of the special orthogonal functions [35].
For a given potential, the Schr\"{o}dinger or Schr\"{o}dinger-like equations
in spherical coordinates are reduced to a generalized equation of
hypergeometric type with an appropriate coordinate transformation $%
r\rightarrow s$ and then they are solved systematically to find the exact or
particular solutions. The main equation which is closely associated with the
method is given in the following form [27] 
\begin{equation}
\psi ^{\prime \prime }(s)+\frac{\widetilde{\tau }(s)}{\sigma (s)}\psi
^{\prime }(s)+\frac{\widetilde{\sigma }(s)}{\sigma ^{2}(s)}\psi (s)=0,
\end{equation}%
where $\sigma (s)$ and $\widetilde{\sigma }(s)$ are polynomials at most
second-degree, $\widetilde{\tau }(s)$ is a first-degree polynomial and $\psi
(s)$ is a function of the hypergeometric type.

Let us now try to reduce Eq.(1) to a comprehensible form by taking $\psi
(s)=\phi (s)y(s)$ and choosing an appropriate function $\phi (s)$: 
\begin{equation}
y^{\prime \prime }(s)+\left( 2\frac{\phi ^{\prime }(s)}{\phi (s)}+\frac{%
\widetilde{\tau }(s)}{\sigma (s)}\right) y^{\prime }(s)+\left( \frac{\phi
^{\prime \prime }(s)}{\phi (s)}+\frac{\phi ^{\prime }(s)}{\phi (s)}\frac{%
\widetilde{\tau }(s)}{\sigma (s)}+\frac{\widetilde{\sigma }(s)}{\sigma
^{2}(s)}\right) y(s)=0.
\end{equation}%
At the first stage, Eq.(2) can be seen to be more complicated than the main
equation, Eq.(1). To ensure the reasonable understanding, the coefficient of 
$y^{\prime }(s)$ is taken in the form $\tau (s)/\sigma (s)$, where $\tau (s)$
is a polynomial of degree at most one, \textit{i.e.}, 
\begin{equation}
2\frac{\phi ^{\prime }(s)}{\phi (s)}+\frac{\widetilde{\tau }(s)}{\sigma (s)}=%
\frac{\tau (s)}{\sigma (s)},
\end{equation}%
and hence the most regular form is obtained as follows, 
\begin{equation}
\frac{\phi ^{\prime }(s)}{\phi (s)}=\frac{\pi (s)}{\sigma (s)},
\end{equation}%
where 
\begin{equation}
\pi (s)=\frac{1}{2}[\tau (s)-\widetilde{\tau }(s)].
\end{equation}%
The most useful demonstration of Eq. (5) is 
\begin{equation}
\tau (s)=\widetilde{\tau }(s)+2\pi (s).
\end{equation}%
The new parameter $\pi (s)$ is a polynomial of degree at most one. In
addition, the term $\phi ^{\prime \prime }(s)/\phi (s)$ which appears in the
coefficient of $y(s)$ in Eq.(2) is arranged as follows 
\begin{equation}
\frac{\phi ^{\prime \prime }(s)}{\phi (s)}=\left( \frac{\phi ^{\prime }(s)}{%
\phi (s)}\right) ^{\prime }+\left( \frac{\phi ^{\prime }(s)}{\phi (s)}%
\right) ^{2}=\left( \frac{\pi (s)}{\sigma (s)}\right) ^{\prime }+\left( 
\frac{\pi (s)}{\sigma (s)}\right) ^{2}.
\end{equation}%
In this case, the coefficient of $y(s)$ is transformed into a more suitable
arrangement by taking the form in Eq.(4); 
\begin{equation}
\frac{\phi ^{\prime \prime }(s)}{\phi (s)}+\frac{\phi ^{\prime }(s)}{\phi (s)%
}\frac{\widetilde{\tau }(s)}{\sigma (s)}+\frac{\widetilde{\sigma }(s)}{%
\sigma ^{2}(s)}=\frac{\bar{\sigma}(s)}{\sigma ^{2}(s)}
\end{equation}%
where 
\begin{equation}
\bar{\sigma}(s)=\widetilde{\sigma }(s)+\pi ^{2}(s)+\pi (s)[\widetilde{\tau }%
(s)-\sigma ^{\prime }(s)]+\pi ^{\prime }(s)\sigma (s).
\end{equation}%
Substituting the right-hand sides of Eq.(3) and Eq.(8) into Eq.(2), an
equation of the same type as Eq.(1) is obtained as 
\begin{equation}
y^{\prime \prime }(s)+\frac{\tau (s)}{\sigma (s)}y^{\prime }(s)+\frac{\bar{%
\sigma}(s)}{\sigma ^{2}(s)}y(s)=0.
\end{equation}%
As a consequence of the above algebraic transformations, the functional form
of Eq.(1) is protected by following a systematic way. Therefore, the
transformations allow us to replace the function of the hypergeometric type $%
\psi (s)$ by the substitution $\phi (s)y(s)$, where $\phi (s)$ satisfies
Eq.(4) whit an arbitrary linear polynomial $\pi (s)$. If the polynomial $%
\bar{\sigma}(s)$ in Eq.(10) is divisible by $\sigma (s)$, \textit{i.e.}, 
\begin{equation}
\bar{\sigma}(s)=\lambda \sigma (s),
\end{equation}%
where $\lambda $ is a constant, Eq.(10) is reduced to an equation of
hypergeometric type 
\begin{equation}
\sigma (s)y^{\prime \prime }+\tau (s)y^{\prime }+\lambda y=0,
\end{equation}%
and also its solution is given as a function of hypergeometric type [35]. To
determine the polynomial $\pi (s)$, Eq.(9) is compared with Eq.(11) and then
a quadratic equation for $\pi (s)$ is obtained as follows, 
\begin{equation}
\pi ^{2}(s)+\pi (s)[\widetilde{\tau }(s)-\sigma ^{\prime }(s)]+\widetilde{%
\sigma }(s)-k\sigma (s),
\end{equation}%
where 
\begin{equation}
k=\lambda -\pi ^{\prime }(s).
\end{equation}%
The solution of this quadratic equation for $\pi (s)$ yields the following
equality 
\begin{equation}
\pi (s)=\frac{\sigma ^{\prime }(s)-\widetilde{\tau }(s)}{2}\pm \sqrt{\left( 
\frac{\sigma ^{\prime }(s)-\widetilde{\tau }(s)}{2}\right) ^{2}-\widetilde{%
\sigma }(s)+k{\sigma }(s)}.
\end{equation}%
In order to obtain the possible solutions according to the plus and minus
signs of Eq.(15), the parameter $k$ within the square root sign must be
known explicitly. To provide this requirement, the expression under the
square root sign has to be the square of a polynomial, since $\pi (s)$ is a
polynomial of degree at most one. In this case, an equation of the quadratic
form is available for the constant $k$. Setting the discriminant of this
quadratic equal to zero, the constant $k$ is determined clearly. After
determining $k$, the polynomial $\pi (s)$ is obtained from Eq.(15), and then 
$\tau (s)$ and $\lambda $ are also obtained by using Eq.(5) and Eq.(14),
respectively.

A common trend which is followed to generalize the solutions of Eq.(12) is
to show that all the derivatives of functions of hypergeometric type are
also of hypergeometric type. For this purpose, Eq.(12) is differentiated by
using the representation $v_{1}(s)=y^{\prime }(s)$ 
\begin{equation}
\sigma (s)v_{1}^{\prime {\prime }}(s)+\tau _{1}(s)v_{1}^{\prime }(s)+\mu
_{1}v_{1}(s)=0,
\end{equation}%
where $\tau _{1}(s)=\tau (s)+\sigma ^{\prime }(s)$ and $\mu _{1}=\lambda
+\tau ^{\prime }(s)$. $\tau _{1}(s)$ is a polynomial of degree at most one
and $\mu _{1}$ is independent of the variable $s$. It is clear that Eq.(16)
is an equation of hypergeometric type again. By taking $v_{2}(s)=y^{\prime
\prime }(s)$ as a new representation, the second derivation of Eq.(12)
becomes 
\begin{equation}
\sigma (s)v_{2}^{\prime {\prime }}(s)+\tau _{2}(s)v_{2}^{\prime }(s)+\mu
_{2}v_{2}(s)=0,
\end{equation}%
where 
\begin{equation}
\tau _{2}(s)=\tau _{1}(s)+\sigma ^{\prime }(s)=\tau (s)+2\sigma ^{\prime
}(s),
\end{equation}%
\begin{equation}
\mu _{2}=\mu _{1}+\tau _{1}^{\prime }(s)=\lambda +2\tau ^{\prime }(s)+\sigma
^{\prime \prime }(s).
\end{equation}%
In a similar way, an equation of hypergeometric type for $%
v_{n}(s)=y^{(n)}(s) $ is constructed as a family of particular solutions of
Eq.(12) corresponding to a given $\lambda $; 
\begin{equation}
\sigma (s)v_{n}^{\prime {\prime }}(s)+\tau _{n}(s)v_{n}^{\prime }(s)+\mu
_{n}v_{n}(s)=0,
\end{equation}%
and here the general recurrence relations for $\tau _{n}(s)$ and $\mu _{n}$
are found as follows, respectively, 
\begin{equation}
\tau _{n}(s)=\tau (s)+n\sigma ^{\prime }(s),
\end{equation}%
\begin{equation}
\mu _{n}=\lambda +n\tau ^{\prime }(s)+\frac{n(n-1)}{2}\sigma ^{\prime \prime
}(s).
\end{equation}%
When $\mu _{n}=0$, Eq.(22) becomes as follows 
\begin{equation}
\lambda =\lambda _{n}=-n\tau ^{\prime }(s)-\frac{n(n-1)}{2}\sigma ^{\prime
\prime }(s),\quad (n=0,1,2,\ldots )
\end{equation}%
and then Eq.(20) has a particular solution of the form%
\begin{equation*}
y(s)=y_{n}(s)=\frac{B_{n}}{\rho (s)}\frac{d^{n}}{dr^{n}}\left[ \sigma
^{n}(s)\rho (s)\right] ,
\end{equation*}
which is the Rodrigues relation of degree $n$ and $\rho (s)$ is the weight
function satisfying the differential equation%
\begin{equation*}
\left[ \sigma (r)\rho (r)\right] ^{\prime }=\tau (r)\rho (r).
\end{equation*}%
To obtain an eigenvalue solution through the NU method, the relationship
between $\lambda $ and $\lambda _{n}$ must be set up by means of Eq.(14) and
Eq.(23).

\section{Dirac Equation}

In the relativistic description, the Dirac equation of a single-nucleon with
the mass $\mu $ moving in an attractive scalar potential $S(\vec{r})$ and a
repulsive vector potential $V(\vec{r})$ can be written as 
\begin{equation}
\left[ \vec{\alpha}.c\vec{P}+\beta (\mu c^{2}+S(\vec{r}))+V(\vec{r})\right]
\psi _{n_{r}\kappa }(\vec{r})=E_{n_{r}\kappa }\psi _{n_{r}\kappa }(\vec{r}),
\end{equation}%
where 
\begin{equation}
\vec{P}=-i\hbar \vec{\nabla},~~~~~~~~\vec{\alpha}=\left( 
\begin{array}{cc}
0 & \vec{\sigma} \\ 
\vec{\sigma} & 0%
\end{array}%
\right) ,~~~~~~~~~~\beta =\left( 
\begin{array}{cc}
0 & I \\ 
-I & 0%
\end{array}%
\right) ,
\end{equation}%
with $\vec{\sigma}$ is the vector Pauli spin matrix and $I$ is the identity
matrix. $\vec{P}$ is the three momentum operators, $\vec{\alpha}$ and $\beta 
$ are the usual $4\times 4$ Dirac matrices [36], $c$ is the velocity of
light in vacuum and $\hbar $ is the Planck's constant divided by $2\pi $. $%
E_{n_{r}\kappa }$ denotes the relativistic energy eigenvalues of the Dirac
particle. For nuclei with spherical symmetry, $S(\vec{r})$ and $V(\vec{r})$
potentials in Eq.(24) represent only the radial coordinates, \textit{i.e.}, $%
S(\vec{r})=S(r)$ and $V(\vec{r})=V(r)$, where $r$ is the magnitude of $\vec{r%
}$. The spinor wave functions $\psi _{n_{r}\kappa }(\vec{r})$ can be written
in the following form 
\begin{equation}
\psi _{n_{r}\kappa }(\vec{r})=\frac{1}{r}\left( 
\begin{array}{c}
F_{n_{r}\kappa }(r)\left[ Y_{\ell }(\theta ,\phi )\chi _{\pm }\right]
_{m}^{(j)} \\ 
iG_{n_{r}\kappa }(r)\left[ Y_{\tilde{\ell}}(\theta ,\phi )\chi _{\pm }\right]
_{m}^{(j)}%
\end{array}%
\right) ,
\end{equation}%
where $Y_{\ell }(\theta ,\phi )$ ($Y_{\tilde{\ell}}(\theta ,\phi )$) and $%
\chi _{\pm }$ are the spin (pseudospin) spherical harmonic and spin wave
function which are coupled to angular momentum $j$ with projection $m$,
respectively. $F_{n_{r}\kappa }(r)$ and $G_{n_{r}\kappa }(r)$ are the radial
wave functions for the upper and lower components, respectively. The label $%
\kappa $ has two explanations; the aligned spin $j=\ell +1/2$ ($%
s_{1/2},p_{3/2},etc.$) is valid for the case of $\kappa =-(j+1/2)$ and then $%
\tilde{\ell}=\ell +1$, while the unaligned spin $j=\ell -1/2$ ($%
p_{1/2},d_{3/2},etc.$) is valid for the case of $\kappa =(j+1/2)$ and then $%
\tilde{\ell}=\ell -1$. Thus, the quantum number $\kappa $ and the radial
quantum number $n_{r}$ are sufficient to label the Dirac eigenstates. The
Dirac equation given in Eq.(24) may be reduced to a set of two coupled
ordinary differential equations (in units of $c=\hbar =1$): 
\begin{equation}
\left( \frac{d}{dr}+\frac{\kappa }{r}\right) F_{n_{r}\kappa }(r)=(\mu
+E_{n_{r}\kappa }-\Delta (r))G_{n_{r}\kappa }(r),
\end{equation}%
\begin{equation}
\left( \frac{d}{dr}-\frac{\kappa }{r}\right) G_{n_{r}\kappa }(r)=(\mu
-E_{n_{r}\kappa }+\Sigma (r))F_{n_{r}\kappa }(r),
\end{equation}%
where $\Delta (r)=V(r)-S(r)$ and $\Sigma (r)=V(r)+S(r)$ are the difference
and the sum potentials, respectively. By substituting%
\begin{equation*}
F_{n_{r}\kappa }(r)=\frac{1}{(\mu -E_{n_{r}\kappa }+\Sigma (r))}\left( \frac{%
d}{dr}-\frac{\kappa }{r}\right) G_{n_{r}\kappa }(r),
\end{equation*}%
into Eq.(27), the following second order Schr\"{o}dinger-like differential
equation for $G_{n_{r}\kappa }(r)$ can be obtained as%
\begin{equation}
\left( \frac{d^{2}}{dr^{2}}-\frac{\kappa (\kappa -1)}{r^{2}}-(\mu
+E_{n_{r}\kappa }-\Delta (r))(\mu -E_{n_{r}\kappa }+\Sigma (r))-\frac{\frac{%
d\Sigma }{dr}\left( \frac{d}{dr}-\frac{\kappa }{r}\right) }{\mu
-E_{n_{r}\kappa }+\Sigma (r)}\right) G_{n_{r}\kappa }(r)=0,
\end{equation}%
where $E_{n_{r}\kappa }\neq \mu $ when $\Sigma (r)=0$ (exact pseudospin
symmetry). Further, a similar equation for $F_{n_{r}\kappa }(r)$ can be
obtained as follows 
\begin{equation}
\left( \frac{d^{2}}{dr^{2}}-\frac{\kappa (\kappa +1)}{r^{2}}-(\mu
+E_{n_{r}\kappa }-\Delta (r))(\mu -E_{n_{r}\kappa }+\Sigma (r))+\frac{\frac{%
d\Delta }{dr}\left( \frac{d}{dr}+\frac{\kappa }{r}\right) }{\mu
+E_{n_{r}\kappa }-\Delta (r)}\right) F_{n_{r}\kappa }(r)=0,
\end{equation}%
where $E_{n_{r}\kappa }\neq -\mu $ when $\Delta (r)=0$ (exact spin
symmetry). Under the condition of exact spin symmetry, ($d\Delta (r)/dr=0$, 
\textit{i.e.}, $\Delta (r)=C=$constant), Eq. (30) turns out to be 
\begin{equation}
\left( \frac{d^{2}}{dr^{2}}-\frac{\ell \left( \ell +1\right) }{r^{2}}-(\mu
+E_{n_{r}\kappa }-C)\Sigma (r)+E_{n_{r}\kappa }^{2}-\mu ^{2}+C\left( \mu
-E_{n_{r}\kappa }\right) \right) F_{n_{r}\kappa }(r)=0,
\end{equation}%
where $\ell \left( \ell +1\right) $ comes from $\kappa (\kappa +1)$ and $%
\ell \left( \ell +1\right) /r^{2}$ is the spin-centrifugal kinetic energy
term. On the other hand, under the condition of the exact pseudospin
symmetry ($d\Sigma (r)/dr=0$, \textit{i.e.}, $\Sigma (r)=C=$constant), Eq.
(29) is reduced to the form%
\begin{equation}
\left( \frac{d^{2}}{dr^{2}}-\frac{\tilde{\ell}(\tilde{\ell}+1)}{r^{2}}+(\mu
-E_{n_{r}\kappa }+C)\Delta (r)+E_{n_{r}\kappa }^{2}-\mu ^{2}-C\left( \mu
+E_{n_{r}\kappa }\right) \right) G_{n_{r}\kappa }(r)=0,
\end{equation}%
where $\tilde{\ell}(\tilde{\ell}+1)$ comes from $\kappa (\kappa -1)$ and $%
\tilde{\ell}(\tilde{\ell}+1)/r^{2}$ is the pseudo-centrifugal kinetic energy
term. According to the original definition of the pseudo-orbital angular
momentum, the cases $\tilde{\ell}=\kappa -1$ and $\tilde{\ell}=-\kappa $ are
valid for $\kappa >0$ and $\kappa <0$, respectively. Therefore, the
degenerate states come into existence with the same $\tilde{\ell}$ but
different $\kappa $, generating pseudospin symmetry. Another important point
which is necessary to be said on Eq.(32) is that the radial part of the
spinor wave function $\psi _{n_{r}\kappa }(\vec{r})$ must satisfy the
boundary conditions that $G_{n_{r}\kappa }(r)/r$ becomes zero when $%
r\rightarrow \infty $, and $G_{n_{r}\kappa }(r)/r$ is finite at $r=0$.

\section{Bound State Solution by means of the NU Method}

\subsection{Hulth$\mathbf{{\acute{e}}}$n Square Approximation}

In this section, we shall involve the Hulth$\mathbf{{\acute{e}}}$n potential
to solve the Dirac equation given in Eq.(32), meaning that the potential $%
\Delta (r)$ is exponential in $r$ and the pseudo-centrifugal kinetic energy
term is quadratic in $1/r$. The exponential potential in $r$ is the famous
Hulth$\mathbf{{\acute{e}}}$n potential [37,38]; 
\begin{equation}
\Delta (r)=-\Delta _{0}\frac{e^{-\delta r}}{1-e^{-\delta r}},
\end{equation}%
where $\delta $ is the screening parameter which is used for determining the
range of the Hulth$\mathbf{{\acute{e}}}$n potential. The parameter $\Delta
_{0}$ represents $\delta Ze^{2}$, where $Ze$ is the charge of the nucleon
[39]. The intensity of the Hulth$\mathbf{{\acute{e}}}$n potential is denoted
by $\Delta _{0}$ under the condition of $\delta >0$. This potential has been
used in several branches of physics and its discrete and continuum states
have been studied by a variety of techniques such as the algebraic
perturbation calculations which are based upon the dynamical group structure
SO(2,1) [40], the formalism of supersymmetric quantum mechanics within the
framework of the variational method [41], the supersymmetry and shape
invariance property [42], the asymptotic iteration method [43,44] and the
approach proposed by Biedenharn for the Dirac-Coulomb problem [45,46]. With
this potential in place, Eq.(32) has to be solved numerically because the
exponential behavior of $\Delta (r)$ is not compatible with the quadratic
behavior of the pseudo-centrifugal kinetic energy term. However, Eq.(32) is
analytically solvable only for the zero value of the pseudo-orbital angular
momentum, \textit{i.e.}, $\tilde{\ell}=0$ ($\kappa =1$). In order to obtain
more realistic results relating to the degenerate states, the Dirac equation
should be solved for any $\tilde{\ell}$-states. In one of the methods used
for solving Eq.(32), Hulth$\mathbf{{\acute{e}}}$n square approximation can
be introduced as an effective approximation to the pseudo-centrifugal
kinetic energy term in the case of $\tilde{\ell}>0$ and small $r$. Following
the original work of Filho $et~al$ [41] for this approximation, an effective
potential term can be considered as follows 
\begin{equation}
\frac{\tilde{\ell}(\tilde{\ell}+1)\delta ^{2}e^{-2\delta r}}{(1-e^{-\delta
r})^{2}}=\frac{\tilde{\ell}(\tilde{\ell}+1)\delta ^{2}}{([1+\delta
r+...]-1)^{2}}\simeq \frac{\tilde{\ell}(\tilde{\ell}+1)}{r^{2}}.
\end{equation}%
The exponential numerator in Eq.(34) is expanded for small values of $r$ and
higher-order terms are ignored up to first-order term. Recently, the authors
of [42,44,46] have been used a more efficient approximation than that of
Eq.(34) instead of the pseudo-centrifugal kinetic energy term $\tilde{\ell}(%
\tilde{\ell}+1)/r^{2}$. This approximation has the advantage that it is only
valid for small values of $\delta $ and $\tilde{\ell}$. Whereas the present
approximation in Eq.(34) can also be used for small values of $\delta $ and $%
\tilde{\ell}$ ..

When $\Delta (r)$ is taken as the Hulth$\mathbf{{\acute{e}}}$n potential and
the approximation of the centrifugal term as in Eq.(34), Eq.(32) yields%
\begin{equation}
\left[ \frac{d^{2}}{dr^{2}}-\frac{\tilde{\ell}(\tilde{\ell}+1)\delta
^{2}e^{-2\delta r}}{(1-e^{-\delta r})^{2}}-(\mu -E_{n_{r}\kappa }+C)\frac{%
\Delta _{0}e^{-\delta r}}{1-e^{-\delta r}}+E_{n_{r}\kappa }^{2}-\mu
^{2}-C\left( \mu +E_{n_{r}\kappa }\right) \right] G_{n_{r}\kappa }(r)=0,
\end{equation}%
where $\kappa =\tilde{\ell}+1$ for $\kappa >0$ and $\kappa =-\tilde{\ell}$
for $\kappa <0$ and the wave function has to satisfy the boundary
conditions, i.e., $G_{n_{r}\kappa }(r=0)=0$ and $G_{n_{r}\kappa
}(r\rightarrow \infty )=0.$ It is convenient to introduce the following
variable and parameters: 
\begin{equation}
s=e^{-\delta r},\text{ }r\in (0,\infty )\rightarrow \text{s}\in \lbrack 0,1]
\end{equation}%
\begin{equation}
\nu _{1}^{2}=\frac{(\mu -E_{n_{r}\kappa }+C)\Delta _{0}}{\delta ^{2}},
\end{equation}%
\begin{equation}
\omega _{1}^{2}=\frac{E_{n_{r}\kappa }^{2}-\mu ^{2}-C\mu -CE_{n_{r}\kappa }}{%
\delta ^{2}},
\end{equation}%
\begin{equation}
A_{1}=\omega _{1}^{2}+\nu _{1}^{2}-\tilde{\ell}(\tilde{\ell}+1),
\end{equation}%
\begin{equation}
B_{1}=2\omega _{1}^{2}+\nu _{1}^{2},
\end{equation}%
which allows us to rewrite Eq.(35) in the simple form 
\begin{equation}
\left( \frac{d^{2}}{ds^{2}}+\frac{1-s}{s(1-s)}\frac{d}{ds}+\frac{%
A_{1}s^{2}-B_{1}s+\omega _{1}^{2}}{s^{2}(1-s)^{2}}\right) G_{n_{r}\kappa
}(s)=0,
\end{equation}%
where the finiteness of our solution requires that $G_{n_{r}\kappa }(s=1)=0$
for $r\rightarrow 0$ and $G_{n_{r}\kappa }(s=0)=0$ for $r\rightarrow \infty
. $ The above equation can be solved by using a special solution method
mentioned in Ref.[27] and following a short-cut procedure given in Section
2. First of all, before starting the procedure of the solution, Eq.(41) is
compared with the hypergeometric type differential equation given in Eq.(1)
and consequently Eq.(41) is solved analytically due to the fact that the
solution is still subjected to a methodology using algebra and calculus.
This part will be treated in the next subsection partially.

\subsection{Pseudospin Symmetry Solution}

By applying the basic ideas of Ref.[27] and imposing the theory of
orthogonal functions which are known as a generalization of the Rodrigues
formula [35], the comparison of the differential equations in Eq.(41) and
Eq.(1) gives us the following polynomials; 
\begin{equation}
\widetilde{\tau }(s)=1-s,~~~{\sigma }(s)=s(1-s),~~~\widetilde{\sigma }%
(s)=A_{1}s^{2}-B_{1}s+\omega _{1}^{2}.
\end{equation}%
In the present case, if we want to substitute the polynomials given by
Eq.(42) into Eq.(15), the following equality for the polynomial $\pi (s)$ is
obtained 
\begin{equation}
\pi (s)=-\frac{s}{2}\pm \frac{1}{2}\sqrt{(1-4A_{1}-4k)s^{2}+4(B_{1}+k)s-4%
\omega _{1}^{2}}.
\end{equation}%
The expression under the square root of the above equation must be the
square of a polynomial of first degree. This is possible only if its
discriminant is zero and the constant parameter $k$ can be determined from
the condition that the expression under the square root has a double zero.
Hence, $k$ is obtained as $k_{+,-}=2\omega _{1}^{2}-B_{1}\pm i\omega
_{1}\left( 2\tilde{\ell}+1\right) $. In that case, it can be written in the
four possible forms of $\pi (s)$; 
\begin{equation}
\left\{ 
\begin{array}{cc}
\pi (s)=-\frac{s}{2}\pm \frac{1}{2}\left( -\left[ 2\tilde{\ell}+1-2i\omega
_{1}\right] s-2i\omega _{1}\right) , & \text{for }k_{+}=2\omega
_{1}^{2}-B_{1}+i\omega _{1}\left( 2\tilde{\ell}+1\right) , \\ 
\pi (s)=-\frac{s}{2}\pm \frac{1}{2}\left( -\left[ 2\tilde{\ell}+1+2i\omega
_{1}\right] s+2i\omega _{1}\right) , & \text{for }k_{-}=2\omega
_{1}^{2}-B_{1}-i\omega _{1}\left( 2\tilde{\ell}+1\right) .%
\end{array}%
\right.
\end{equation}%
One of the four possible forms of $\pi (s)$ must be chosen to obtain an
eigenvalue equation. Therefore, its most suitable form can be established by%
\begin{equation*}
\pi (s)=i\omega _{1}-\left( i\omega _{1}+\tilde{\ell}+1\right) s,
\end{equation*}%
for $k_{-}$. The trick in this selection is to find the negative derivative
of $\tau (s)$ given in Eq.(6). Hence, $\tau (s)$ and $\tau ^{\prime }(s)$
are obtained as 
\begin{equation}
\tau (s)=1+2i\omega _{1}-(2i\omega _{1}+2\tilde{\ell}+3)s,\text{ }\tau
^{\prime }(s)=-(2i\omega _{1}+2\tilde{\ell}+3)<0~.
\end{equation}%
In this case, a new eigenvalue equation for the Dirac equation becomes 
\begin{equation}
\lambdabar _{n_{r}}=n_{r}^{2}+2n_{r}\left( \tilde{\ell}+1\right)
+2n_{r}i\omega _{1},
\end{equation}%
where it is beneficial to invite the quantity $\lambdabar
_{n_{r}}=-n_{r}\tau ^{\prime }(s)-\frac{n_{r}(n_{r}-1)}{2}\sigma ^{\prime
\prime }(s)$ in Eq.(23). An other eigenvalue equation is obtained from the
equality $\lambdabar =k_{-}+\pi ^{\prime }$ in Eq.(14), 
\begin{equation}
\lambdabar =-\nu ^{2}-\left( \tilde{\ell}+1\right) (1+2i\omega ).
\end{equation}%
In order to find an eigenvalue equation, the right-hand sides of Eq.(46) and
Eq.(47) must be compared with each other. In this case the result obtained
will depend on $E_{n_{r}\kappa }$ in the closed form: 
\begin{equation}
-\omega _{1}^{2}=\left( \frac{(1+2n_{r})\left( \tilde{\ell}+1\right)
+n_{r}^{2}+\nu _{1}^{2}}{2\left( n_{r}+\tilde{\ell}+1\right) }\right) ^{2}.
\end{equation}%
Substituting the terms of right-hand sides of Eqs.(37) and (38) into
Eq.(48), the energy eigenvalue equation for $E_{n_{r}\kappa }$ can be
immediately obtained; 
\begin{equation}
\left( 1+\left( \frac{\Delta _{0}}{Y\delta }\right) ^{2}\right)
E_{n_{r}\kappa }^{2}-\left( C+\frac{2T\Delta _{0}}{Y^{2}}\right)
E_{n_{r}\kappa }+\left( \frac{T\delta }{Y}\right) ^{2}-\mu ^{2}-C\mu =0,
\end{equation}%
where 
\begin{equation}
U=(1+2n_{r})\left( \tilde{\ell}+1\right) +n_{r}^{2},
\end{equation}%
\begin{equation}
Y=2\left( n_{r}+\tilde{\ell}+1\right) ,
\end{equation}%
\begin{equation}
T=U+\frac{(C+\mu )\Delta _{0}}{\delta ^{2}}.
\end{equation}%
The energy spectrum of the Dirac equation for $\Delta (r)=V(r)-S(r)=-\Delta
_{0}\frac{e^{-\delta r}}{1-e^{-\delta r}}$ is obtained by means of Eq.(49).
In this case, the states with the same $n_{r}$ and $\tilde{\ell}$ will be
degenerate. The two energy solutions of the quadratic equation can be
obtained as 
\begin{equation}
E_{n_{r}\kappa }^{\pm }=\frac{\delta ^{2}(2\Delta _{0}T+CY^{2})\pm \sqrt{%
4\delta ^{2}(\Delta _{0}(C+\mu )-\delta ^{2}T)(\Delta _{0}\mu +\delta
^{2}T)Y^{2}+\delta ^{4}(C+2\mu )^{2}Y^{4}}}{2(\Delta _{0}^{2}+Y^{2}\delta
^{2})}.
\end{equation}%
For a given value of $n_{r}$ and $\kappa $ (or $\widetilde{\ell }$), the
above equation provides two distinct positive and negative energy spectra
related with $E_{n_{r}\kappa }^{+}$ or $E_{n_{r}\kappa }^{-}$, respectively.
One of the distinct solutions is only valid to obtain the negative-energy
bound states in the limit of the pseudospin symmetry. Before seeking the
acceptable solution, it is useful to present some analogy about the energy
spectra.

Now, we are going to find the corresponding wave functions for the present
potential model. Firstly, we calculate the weight function defined as [47-51]%
\begin{equation}
\rho (s)=\frac{1}{\sigma (s)}\exp \left( \int \frac{\tau (s)}{\sigma (s)}%
ds\right) =s^{2i\omega _{1}}\left( 1-s\right) ^{2\tilde{\ell}+1},
\end{equation}%
and the first part of the wave function in Eq. (4):%
\begin{equation}
\phi (s)=\exp \left( \int \frac{\pi (s)}{\sigma (s)}ds\right) =s^{i\omega
_{1}}\left( 1-s\right) ^{\tilde{\ell}+1}.
\end{equation}%
Hence, the second part of the wave function which is the solution of Eq.(20)
can be obtained by means of the so called Rodrigues representation 
\begin{equation*}
y_{n_{r}}(s)=c_{n_{r}\kappa }s^{-2i\omega _{1}}\left( 1-s\right) ^{-\left( 2%
\tilde{\ell}+1\right) }\frac{d^{n_{r}}}{ds^{n_{r}}}\left[ s^{n_{r}+2i\omega
_{1}}\left( 1-s\right) ^{n_{r}+2\tilde{\ell}+1}\right]
\end{equation*}%
\begin{equation}
\sim P_{n_{r}}^{\left( 2i\omega _{1},2\tilde{\ell}+1\right) }(1-2s),\text{ }%
s\in \lbrack 0,1],
\end{equation}%
where the Jacobi polynomial $P_{n_{r}}^{\left( \mu ,\nu \right) }(x)$ is
defined for $\func{Re}$($\nu )>-1$ and $\func{Re}$($\mu )>-1$ for the
argument $x\in \left[ -1,+1\right] $ and $c_{n_{r}\kappa }$ is the
normalization constant$.$ By using $G_{n_{r}\kappa }(s)=\phi
(s)y_{n_{r}}(s), $ in this way we may write the lower-spinor wave function
in the following fashion%
\begin{equation*}
G_{n_{r}\kappa }(r)=c_{n_{r}\kappa }\left( \exp (-i\omega _{1}\delta
r)\right) \left( 1-\exp (-\delta r)\right) ^{\tilde{\ell}+1}P_{n}^{\left(
2i\omega _{1},2\tilde{\ell}+1\right) }(1-2\exp (-\delta r))
\end{equation*}%
\begin{equation*}
=c_{n_{r}\kappa }\frac{\left( 2i\omega _{1}+1\right) _{n_{r}}}{n_{r}!}\left(
\exp (-i\omega _{1}\delta r)\right) \left( 1-\exp (-\delta r)\right) ^{%
\tilde{\ell}+1}
\end{equation*}%
\begin{equation}
\times 
\begin{array}{c}
_{2}F_{1}%
\end{array}%
\left( -n_{r},n_{r}+2\left( i\omega _{1}+\tilde{\ell}+1\right) ;1+2i\omega
_{1};\exp (-\delta r)\right) ,
\end{equation}%
where%
\begin{equation}
i\omega _{1}\delta =\sqrt{C\left( \mu +E_{n_{r}\kappa }\right) +\mu
^{2}-E_{n_{r}\kappa }^{2}}>0.
\end{equation}%
The hypergeometric series $%
\begin{array}{c}
_{2}F_{1}%
\end{array}%
\left( -n_{r},n_{r}+2\left( i\omega _{1}+\tilde{\ell}+1\right) ;1+2i\omega
_{1};\exp (-\delta r)\right) $ terminates for $n_{r}=0$ and thus converges
for all values of real parameters $\omega _{1}>0$ and $\tilde{\ell}>0.$ In
case if $C=0,$ then $i\omega _{1}\delta =\sqrt{\left( \mu +E_{n_{r}\kappa
}\right) \left( \mu -E_{n_{r}\kappa }\right) }$ with the following
restriction $E_{n_{r}\kappa }<\mu $ required to obtain bound state (real)
solutions for both positive and negative solutions of $E_{n_{r}\kappa }$ in
Eq. (53)$.$ Now, before presenting the corresponding upper-component $%
F_{n_{r}\kappa }(r),$ let us recall a recurrence relation of hypergeometric
function%
\begin{equation}
\frac{d}{ds}\left[ 
\begin{array}{c}
_{2}F_{1}%
\end{array}%
\left( a;b;c;s\right) \right] =\left( \frac{ab}{c}\right) 
\begin{array}{c}
_{2}F_{1}%
\end{array}%
\left( a+1;b+1;c+1;s\right) ,
\end{equation}%
with which the corresponding upper component $F_{n_{r}\kappa }(r)$ can be
given by solving Eq. (28) as follows 
\begin{equation*}
F_{n_{r}\kappa }(r)=d_{n_{r}\kappa }\frac{\left( \exp (-i\omega _{1}\delta
r)\right) \left( 1-\exp (-\delta r)\right) ^{\tilde{\ell}+1}}{(\mu
-E_{n_{r}\kappa }+C)}\left[ \frac{\left( \tilde{\ell}+1\right) \delta \exp
(-\delta r)}{\left( 1-\exp (-\delta r)\right) }-i\omega _{1}\delta -\frac{%
\kappa }{r}\right]
\end{equation*}%
\begin{equation*}
\times 
\begin{array}{c}
_{2}F_{1}%
\end{array}%
\left( -n_{r},n_{r}+2\left( i\omega _{1}+\tilde{\ell}+1\right) ;1+2i\omega
_{1};\exp (-\delta r)\right)
\end{equation*}%
\begin{equation*}
+d_{n_{r}\kappa }\left[ \frac{n_{r}\left[ n_{r}+2\left( i\omega _{1}+\tilde{%
\ell}+1\right) \right] \delta \left( \exp (-\delta r)\right) ^{i\omega
_{1}+1}\left( 1-\exp (-\delta r)\right) ^{\tilde{\ell}+1}}{\left( 1+2i\omega
_{1}\right) (\mu -E_{n_{r}\kappa }+C)}\right]
\end{equation*}%
\begin{equation}
\times 
\begin{array}{c}
_{2}F_{1}%
\end{array}%
\left( 1-n_{r},n_{r}+2\left( i\omega _{1}+\tilde{\ell}+\frac{3}{2}\right)
;2\left( 1+i\omega _{1}\right) ;\exp (-\delta r)\right) ,
\end{equation}%
where $E_{n_{r}\kappa }\neq \mu $ when $C=0$, exact pseudospin symmetry and $%
d_{n_{r}\kappa }$ is the normalization factor.

\subsection{Spin Symmetry Solution}

This symmetry arises from the near equality in magnitude of an attractive
scalar, $S(\vec{r}),$ and repulsive vector, $V(\vec{r}),$ relativistic mean
field, $S(\vec{r})\sim V(\vec{r})$ in which the nucleon move [48-51].
Therefore, we simply take the sum potential equal to the isotonic potential
model, i.e.,%
\begin{equation}
\Sigma (r)=-\Sigma _{0}\frac{e^{-\delta r}}{1-e^{-\delta r}}.
\end{equation}%
along with the approximation given by Eq.(34) to deal with the spin-orbit
centrifugal term $\ell (\ell +1)/r^{2}$. In the last equation, the choice of 
$\Sigma (r)=2V(r)\rightarrow V(r)$ allows us to reduce the resulting
solutions of the Dirac equation into their non-relativistic limits under
appropriate choice of parameter transformations [51]. Therefore, the
spin-symmetry Dirac equation (31) becomes%
\begin{equation}
\left\{ \frac{d^{2}}{dr^{2}}-\frac{\ell (\ell +1)\delta ^{2}e^{-2\delta r}}{%
(1-e^{-\delta r})^{2}}+(\mu +E_{n_{r}\kappa }-C)\frac{\Sigma _{0}e^{-\delta
r}}{1-e^{-\delta r}}-\left[ \mu ^{2}-E_{n_{r}\kappa }^{2}-C\left( \mu
-E_{n_{r}\kappa }\right) \right] \right\} F_{n_{r}\kappa }(r)=0,
\end{equation}%
where $\kappa =\ell $ for $\kappa >0$ and $\kappa =-(\ell +1)$ for $\kappa
<0.$ It is convenient to introduce the following new variable and
parameters: 
\begin{equation}
s=e^{-\delta r},\text{ }r\in (0,\infty )\rightarrow \text{s}\in \lbrack 0,1]
\end{equation}%
\begin{equation}
\nu _{2}^{2}=\frac{(\mu +E_{n_{r}\kappa }-C)\Sigma _{0}}{\delta ^{2}},
\end{equation}%
\begin{equation}
\omega _{2}^{2}=\frac{E_{n_{r}\kappa }^{2}-\mu ^{2}+C\mu -CE_{n_{r}\kappa }}{%
\delta ^{2}},
\end{equation}%
\begin{equation}
A_{2}=\omega _{2}^{2}-\nu _{2}^{2}-\ell (\ell +1),
\end{equation}%
\begin{equation}
B_{2}=2\omega _{2}^{2}-\nu _{2}^{2},
\end{equation}%
which allow us to rewrite Eq.(62) in a more simple form as 
\begin{equation}
\left( \frac{d^{2}}{ds^{2}}+\frac{1-s}{s(1-s)}\frac{d}{ds}+\frac{%
A_{2}s^{2}-B_{2}s+\omega _{2}^{2}}{s^{2}(1-s)^{2}}\right) F_{n_{r}\kappa
}(s)=0,
\end{equation}%
where the finiteness of our solutions require that $F_{n_{r}\kappa }(1)=0$
and $F_{n_{r}\kappa }(0)\rightarrow 0.$ We apply the NU method following the
same steps of solution in previous section to obtain the expressions: 
\begin{equation}
\widetilde{\tau }(s)=1-s,\text{\ }\sigma (s)=s(1-s),\text{\ }\widetilde{%
\sigma }(s)=A_{2}s^{2}-B_{2}s+\omega _{2}^{2}.
\end{equation}%
To avoid repition, the functions required by the method for $\pi (s),$ $k$
and $\tau (s)$ can be established as 
\begin{equation}
\pi (s)=i\omega _{2}-\left( i\omega _{2}+\ell +1\right) s,
\end{equation}%
\begin{equation}
k=2\omega _{2}^{2}-B_{2}-i\omega _{2}\left( 2\ell +1\right) ,
\end{equation}%
and%
\begin{equation}
\tau (s)=1+2i\omega _{2}-(2i\omega _{2}+2\ell +3)s,\text{ }\tau ^{\prime
}(s)=-(2i\omega _{2}+2\ell +3)<0~.
\end{equation}%
respectively, with prime denotes the derivative with respect to $s.$ Also,
the parameters $\lambdabar $ and $\lambdabar _{n}$ take the forms:%
\begin{equation}
\lambdabar _{n_{r}}=n_{r}^{2}+2n_{r}\left( \ell +1\right) +2n_{r}i\omega _{2}%
\text{ and }\lambdabar =\nu _{2}^{2}-\left( \ell +1\right) (1+2i\omega _{2}).
\end{equation}%
Using the basic condition $\lambdabar =\lambdabar _{n}$ followed by simple
algebra$,$ we obtain 
\begin{equation}
-\omega _{2}^{2}=\left( \frac{(1+2n_{r})\left( \ell +1\right) +n_{r}^{2}-\nu
_{2}^{2}}{2\left( n_{r}+\ell +1\right) }\right) ^{2},\text{ }n_{r},\ell
=1,2,3,\cdots 
\end{equation}%
and then the energy eigenvalue equation is immediately obtained%
\begin{equation}
\left( 1+\left( \frac{\Sigma _{0}}{Z\delta }\right) ^{2}\right)
E_{n_{r}\kappa }^{2}-\left( C+\frac{2S\Sigma _{0}}{Z^{2}}\right)
E_{n_{r}\kappa }+\left( \frac{S\delta }{Z}\right) ^{2}-\mu ^{2}+C\mu =0,
\end{equation}%
where 
\begin{equation}
W=(1+2n_{r})\left( \ell +1\right) +n_{r}^{2},
\end{equation}%
\begin{equation}
Z=2\left( n_{r}+\ell +1\right) ,
\end{equation}%
\begin{equation}
S=W+\frac{(C-\mu )\Sigma _{0}}{\delta ^{2}}.
\end{equation}%
The two energy solutions of the quadratic equation (75) can be obtained as 
\begin{equation}
E_{n_{r}\kappa }^{\pm }=\frac{\delta ^{2}(2\Sigma _{0}S+CZ^{2})\pm \sqrt{%
4\delta ^{2}(\Sigma _{0}(C-\mu )-\delta ^{2}S)(-\Sigma _{0}\mu +\delta
^{2}S)Z^{2}+\delta ^{4}(C-2\mu )^{2}Z^{4}}}{2(\Sigma _{0}^{2}+Z^{2}\delta
^{2})}.
\end{equation}%
For a given value of $n_{r}$ and $\kappa $ (or $\ell $), the above equation
provides two distinct positive and negative energy spectra related with $%
E_{n_{r}\kappa }^{+}$ or $E_{n_{r}\kappa }^{-}$, respectively. One of the
distinct solutions is only valid to obtain the positive-energy bound states
in the limit of the spin symmetry.

In our calculations for the spin symmetry wave functions, we firstly find
the weight function:%
\begin{equation}
\rho (s)=s^{2i\omega _{2}}\left( 1-s\right) ^{2\ell +1},
\end{equation}%
and from which the second part of the wave function by means of Rodrigues
formula as 
\begin{equation*}
y_{n_{r}}(s)=a_{n_{r}\kappa }s^{-2i\omega _{2}}\left( 1-s\right) ^{-\left(
2\ell +1\right) }\frac{d^{n_{r}}}{ds^{n_{r}}}\left[ s^{n_{r}+2i\omega
_{2}}\left( 1-s\right) ^{n_{r}+2\ell +1}\right]
\end{equation*}%
\begin{equation}
\sim P_{n_{r}}^{\left( 2i\omega _{2},2\ell +1\right) }(1-2s),\text{ }s\in
\lbrack 0,1],
\end{equation}%
where the Jacobi polynomial $P_{n_{r}}^{\left( \mu ,\nu \right) }(x)$ is
defined for $\func{Re}$($\nu )>-1$ and $\func{Re}$($\mu )>-1$ for the
argument $x\in \left[ -1,+1\right] $ and $a_{n_{r}\kappa }$ is the
normalization constant$.$ Further, the first part of the wave function is
being calculated as%
\begin{equation}
\phi (s)=s^{i\omega _{2}}\left( 1-s\right) ^{\ell +1}.
\end{equation}%
By using $F_{n_{r}\kappa }(s)=\phi (s)y_{n_{r}}(s),$ in the spin symmetry
case, we may write down the upper-spinor wave function in the following
fashion%
\begin{equation*}
F_{n_{r}\kappa }(r)=a_{n_{r}\kappa }\left( \exp (-i\omega _{2}\delta
r)\right) \left( 1-\exp (-\delta r)\right) ^{\ell +1}P_{n_{r}}^{\left(
2i\omega _{2},2\ell +1\right) }(1-2\exp (-\delta r))
\end{equation*}%
\begin{equation*}
=a_{n_{r}\kappa }\frac{\left( 2i\omega _{2}+1\right) _{n_{r}}}{n_{r}!}\left(
\exp (-i\omega _{2}\delta r)\right) \left( 1-\exp (-\delta r)\right) ^{\ell
+1}
\end{equation*}%
\begin{equation}
\times 
\begin{array}{c}
_{2}F_{1}%
\end{array}%
\left( -n_{r},n_{r}+2\left( i\omega _{2}+\ell +1\right) ;1+2i\omega
_{2};\exp (-\delta r)\right) ,
\end{equation}%
where%
\begin{equation}
i\omega _{2}\delta =\sqrt{C\left( E_{n_{r}\kappa }-\mu \right) +\mu
^{2}-E_{n_{r}\kappa }^{2}}>0.
\end{equation}%
It is noted that the hypergeometric series $%
\begin{array}{c}
_{2}F_{1}%
\end{array}%
\left( -n_{r},n_{r}+2\left( i\omega _{2}+\ell +1\right) ;1+2i\omega
_{2};\exp (-\delta r)\right) $ terminates for $n_{r}=0$ and thus it
converges for all values of real parameters $\omega _{2}>0$ and $\ell >0.$
In case when $C=0,$ then $i\omega _{1}\delta =\sqrt{\left( \mu
+E_{n_{r}\kappa }\right) \left( \mu -E_{n_{r}\kappa }\right) }$ with a
restriction for real bound states that $E_{n_{r}\kappa }<\mu $ for both
positive and negative solutions of $E_{n_{r}\kappa }$ in Eq. (79)$.$ Thus,
the corresponding spin-symmetric lower-component $G_{n_{r}\kappa }(r)$ can
be found as follows 
\begin{equation*}
G_{n_{r}\kappa }(r)=b_{n_{r}\kappa }\frac{\left( \exp (-i\omega _{2}\delta
r)\right) \left( 1-\exp (-\delta r)\right) ^{\ell +1}}{(\mu +E_{n_{r}\kappa
}-C)}\left[ \frac{\left( \ell +1\right) \delta \exp (-\delta r)}{\left(
1-\exp (-\delta r)\right) }-i\omega _{2}\delta +\frac{\kappa }{r}\right]
\end{equation*}%
\begin{equation*}
\times 
\begin{array}{c}
_{2}F_{1}%
\end{array}%
\left( -n_{r},n_{r}+2\left( i\omega _{2}+\ell +1\right) ;1+2i\omega
_{2};\exp (-\delta r)\right)
\end{equation*}%
\begin{equation*}
+b_{n_{r}\kappa }\left[ \frac{n_{r}\delta \left[ n_{r}+2\left( \ell
+1+i\omega _{2}\right) \right] \left( \exp (-\delta r)\right) ^{i\omega
_{2}+1}\left( 1-\exp (-\delta r)\right) ^{\ell +1}}{\left( 1+2i\omega
_{1}\right) (\mu +E_{n_{r}\kappa }-C)}\right]
\end{equation*}%
\begin{equation}
\times 
\begin{array}{c}
_{2}F_{1}%
\end{array}%
\left( 1-n_{r},n_{r}+2\left( i\omega _{2}+\ell +\frac{3}{2}\right) ;2\left(
1+i\omega _{2}\right) ;\exp (-\delta r)\right) ,
\end{equation}%
where $E_{n_{r}\kappa }\neq -\mu $ when $C=0$, exact spin symmetry and $%
b_{n_{r}\kappa }$ is the normalization constant.

Let us finally remark that a careful inspection to our present
spin-symmetric solution shows that it can can be easily recovered by knowing
the relationship between the present set of parameters $(\omega _{2}^{2},\nu
_{2}^{2},A_{2},B_{2})$ and the previous set of parameters $(\omega
_{1}^{2},\nu _{1}^{2},A_{1},B_{1}).$ This tells us that the positive energy
solution for spin symmetry (negative energy solution for pseudospin
symmetry) can be obtained directly from those of the negative energy
solution for pseudospin symmetry (positive energy solution for spin
symmetry) by performing the following replacements [48-51]: 
\begin{equation*}
F_{n_{r}\kappa }(r)\leftrightarrow G_{n_{r}\kappa }(r),\text{ }%
V(r)\rightarrow -V(r)\text{ (or }\Sigma _{0}\leftrightarrow -\Delta _{0}%
\text{)},\text{ }\ell (\ell +1)\leftrightarrow \tilde{\ell}(\tilde{\ell}+1)
\end{equation*}%
\begin{equation}
,\text{ }E_{n_{r}\kappa }^{+}\leftrightarrow -E_{n_{r}\kappa }^{-},\text{ }%
\omega _{2}^{2}\leftrightarrow \omega _{1}^{2}\text{ and }\nu
_{2}^{2}\leftrightarrow -\nu _{1}^{2}.
\end{equation}%
That is, with the above replacements, Eqs. (49) and (57) yield Eqs. (75) and
(83) and vice versa is true.

Let us now present the non-relativistic limit. This can be achieved when we
set $C=0,$ $\Sigma _{0}=\delta $ and using the mapping $E_{n_{r}\kappa }-\mu
\rightarrow E_{n_{r}\ell }$ and \ $E_{n_{r}\kappa }+\mu \rightarrow 2\mu $
in Eqs.(64), (65) and (74)$,$ then the resulting energy eigenvalues (in $%
\hbar =c=e=1$ units) are%
\begin{equation}
E_{n_{r}\ell }=-\frac{1}{2\mu }\left[ \frac{(1+2n_{r})\left( \ell +1\right)
\delta +n_{r}^{2}\delta -2\mu }{2\left( n_{r}+\ell +1\right) }\right] ^{2},%
\text{ }n_{r},\ell =0,1,2,3,\cdots .
\end{equation}%
Also, the wave functions in Eqs.(83) and (84) turns out to become%
\begin{equation*}
R_{n_{r}\ell }(r)=a_{n_{r}\ell }r^{-1}\left( \exp (-\sqrt{-2\mu E_{n_{r}\ell
}}r)\right) \left( 1-\exp (-\delta r)\right) ^{\ell +1}P_{n_{r}}^{\left( 2%
\sqrt{-2\mu E_{n_{r}\ell }}/\delta ,2\ell +1\right) }(1-2\exp (-\delta r))
\end{equation*}%
\begin{equation*}
=a_{n_{r}\ell }\frac{\left( 2\sqrt{-2\mu E_{n_{r}\ell }}/\delta +1\right)
_{n_{r}}}{n_{r}!}r^{-1}\left( \exp (-\sqrt{-2\mu E_{n_{r}\ell }}r)\right)
\left( 1-\exp (-\delta r)\right) ^{\ell +1}
\end{equation*}%
\begin{equation}
\times 
\begin{array}{c}
_{2}F_{1}%
\end{array}%
\left( -n_{r},n_{r}+2\left( \sqrt{-2\mu E_{n_{r}\ell }}/\delta +\ell
+1\right) ;1+2\sqrt{-2\mu E_{n_{r}\ell }}/\delta ;\exp (-\delta r)\right) ,%
\text{ }E_{n_{r}\ell }<0.
\end{equation}

\section{Results and Conclusions}

In the present study, the Dirac equation for the Hulth$\mathbf{{\acute{e}}}$%
n potential is approximately solved under the condition of the exact spin
and pseudospin symmetry within the framework of the relativistic mean field
theory. By using the basic ideas of the NU method, the energy eigenvalue
expression for the arbitrary pseudo-orbital angular momentum $\tilde{\ell}$
is obtained approximately. \ The second-order differential equation given in
Eq.(32) is solved by applying the Hulth$\mathbf{{\acute{e}}}$n square
approximation to deal with the pseudospin--orbit and spin-orbit centrifugal
and kinetic energy terms $\tilde{\ell}(\tilde{\ell}+1)/r^{2}$ and $\ell
(\ell +1)/r^{2}$. The energy spectrum for any $\tilde{\ell}$ states is
obtained analytically. Under the condition of the exact pseudospin and spin
symmetry limitations, the energy relations in the Dirac equation with equal
scalar and vector Hulth$\mathbf{{\acute{e}}}$n potentials are recovered to
see degenerate states.\newline
The results obtained for this motivation show the orbital dependency of the
Dirac equation for the Hulth$\mathbf{{\acute{e}}}$n potential. Certainly, an
analysis detailed by solving Dirac equation in relativistic mean field
theories needs to use a very large scale ($\sim $ 660 MeV) comparing to the
nuclear physics scale (few MeV) ) in point of the intensity of potentials
[6]. For this reason, the intensity of the potential, $\Delta _{0}$, used in
Eq.(33) is considered as 3.4 $fm^{-1}$. The units of $\hbar =c=e=1$ are used
throughout the present work for the sake of simplicity. Hence, the energy
eigenvalue expression given in Eq.(53) can be simply discussed by using a
set of physical parameter values. In the below explanations, although the
energy spectrums can be calculated in dimensionless or arbitrary units, the
calculations are preferably made in $fm^{-1}$ for the energy, mass, $C$ and
intensity of the potential.

It is noted that the energy spectrum given by Eq.(53) indicates a family of
the pseudospin symmetry Hulth$\mathbf{{\acute{e}}}$n potential. Moreover,
the analytical expression for Eq.(53) can be confronted with the results of
[44] which is slightly in agreement with the result presented in Eq.(47) for
the pseudospin symmetry solution. The results are only valid for small
values of $\delta $ and $\tilde{\ell}$ (or $\kappa $). This spectrum changes
with the relevant quantum numbers as well as the screening parameter $\delta 
$. The variation of the energy spectrums ($E_{n_{r}\kappa }^{+}$ and $%
E_{n_{r}\kappa }^{-}$) according to the screening parameter $\delta $ is
shown in Fig.1a and Fig.1b, with the choices of parameters $C=-4.9~fm^{-1}$
and $\mu =5.0~fm^{-1}$, which is in the range of nucleon mass value ($\sim $
1 GeV). Figure 1a indicates the positive-energy bound states, \textit{i.e.}, 
$E_{n_{r}\kappa }^{+}$, while Fig.1b shows the negative-energy bound states, 
\textit{i.e.}, $E_{n_{r}\kappa }^{-}$. For a given value of $n_{r}$ and $%
\widetilde{\ell }$, it is seen that an increment on $\delta $ leads to an
increment on $E_{n_{r}\kappa }^{-}$ along the negative-energy direction
whereas the same increment on $\delta $ results with a reduction on $%
E_{n_{r}\kappa }^{+}$ along the positive-energy direction. The results
presented in Fig.1a show that the energy difference between the states is
still small although the values of screening parameter $\delta $ increases.
Figure 1b has two interesting results: The first one indicates that the
negative-energy bond states appear with the large values of $\delta $ and $%
\tilde{\ell}$. The reason of this aspect comes from the approximation
mentioned in the previous sections. The second one belongs to the small
values of $\delta $. For instance, $E_{n_{r}\kappa }^{-}$ for $1s_{1/2}$, $%
1d_{7/2}$ and $1g_{9/2}$ is valid under the condition of $\delta \gtrsim
0.09 $, $\delta \gtrsim 0.05$ and $\delta \gtrsim 0.03$, respectively.
Therefore, $E_{n_{r}\kappa }^{-}$ still represents the negative-energy bound
states for small values of $\delta $ when $\widetilde{\ell }$ increases.
These results can be also expanded on the other states of the Hulth$\mathbf{{%
\acute{e}}}$n potential with the pseudospin symmetry.

It is well-known that for the finite nuclei the constant $C$ is adjusted to
zero because each potential goes to zero at large distances. If the
difference between the scalar $S(r)$ and vector $V(r)$ potentials equals to
a given constant $C$, this is equivalent to adding the relevant constant to
the relativistic energy and mass. The constant for the energy is unimportant
because it does not affect the energy difference. Whereas the variation of $%
C $ is equivalent to the variation of mass. This is more physically
transparent for the pseudospin-orbit dependency of the Dirac equation.
Moreover, in the case of the infinite nuclear matter, the constant $C$ could
be non-zero. The energy spectrum versus the mass $\mu$ is plotted by setting 
$C =-4.9~fm^{-1}$, as shown in Fig.2. The variation of the energy spectrum
for $\tilde{\ell}=1$, $\tilde{\ell}=3$, $\tilde{\ell}=5$ and $\tilde{\ell}=7$
is presented by using $\delta=0.25$. The radial quantum number is fixed to $%
1 $ ($n_r=1$). In Fig.2, the dashed (red lines, see colour online) and solid
(blue lines, see colour online) lines represent $E_{n_r\kappa}^{+}$ and $%
E_{n_r\kappa}^{-}$, respectively. According to Fig.2, there are two
different regions of energy spectrum versus the mass. For $0<\mu\leq 2.3$ $%
fm^{-1}$ the energy spectrum is in the negative region completely. With the $%
\tilde{\ell}$ increasing, $E_{n_r\kappa}^{+}$ fan out along the positive
part of the energy spectrum whereas $E_{n_r\kappa}^{-}$ overlaps in going
from $\tilde{\ell}=1$ to $\tilde{\ell}=7$. The case of $E_{n_r\kappa}^{-}$
in the interval of $0<\mu\leq 2.3$ $fm^{-1}$ represents the degenerate
states in the different values of $\tilde{\ell}$ for a given value of $n_r$.
A similar trend is seen when $\delta\geq 3.2~fm^{-1}$ in the several values
of $\tilde{\ell}$ for $E_{n_r\kappa}^{+}$ by crossing the zero axis toward
the positive direction of the energy spectrum. However, the numerical
results of $E_{n_r\kappa}^{+}$ over the axis are not relevant for the
negative-energy bound states. Meanwhile, the results of $E_{n_r\kappa}^{-}$
are also valid under the conditions of $\delta\geq
3.9,~3.5,~3.3,~3.2~fm^{-1} $ for $\tilde{\ell}=1,~3,~5,~7$, respectively,
but only relevant in the large values of $\tilde{\ell}$. Furthermore, the
energy spectrum versus the constant $C$ is plotted by taking $\mu =
5.0~fm^{-1}$ and $\delta=0.25$ as shown in Fig.3. According to Fig.3, it is
seen that the negative values of $C $ show more strongly binding energies
under the condition of $C\leq -11.0~fm^{-1}$ for $E_{n_r\kappa}^{+}$ (dashed
lines) and $E_{n_r\kappa}^{-}$ (blue line) in the whole values of $\tilde{%
\ell}$. Moreover, $E_{n_r\kappa}^{+}$ (dashed line) still shows the
negative-energy bound states on condition that $-8<C<-5~fm^{-1}$ up to the
zero axis. In the case of $E_{n_r\kappa}^{-}$ (blue lines) the situation
becomes different than that of $E_{n_r\kappa}^{+}$. The axis is crossed with
changing $\tilde{\ell}$ for $E_{n_r\kappa}^{-}$. The crossing points of axis
from $C=-8~fm^{-1}$ become large with the $\tilde{\ell}$ increasing. In
Table 1, the degenerate states are presented for a Dirac particle within the
Hulth$\mathbf{{\acute{e}}}$n potential, with $C=-4.9~fm^{-1}$, $%
\mu=5.0~fm^{-1}$ and $\Delta_0=3.4~fm^{-1}$. The several pseudo-orbital and
radial quantum numbers are used in the numerical calculations to predict the
orbital dependency of the Dirac equation under the condition of the exact
pseudospin symmetry. As an example, the Dirac eigenstate $1s_{1/2}$ with $%
n_r=1$ and $\kappa=-1$ will have a partner which is denoted by the $0d_{3/2}$
with $n_r-1=0$ and $\kappa=2$. These states are called the pseudospin
partner and degenerated with each other.

\acknowledgments The partial support provided by the Scientific and
Technical Research Council of Turkey (T\"{U}B\.{I}TAK) is highly
appreciated. One of the authors (C.B.) acknowledges the financial support
provided by the Science Foundation of Erciyes University.

\newpage

{\normalsize 
}

\bigskip

\baselineskip= 2\baselineskip
\bigskip \newpage

\bigskip

{\normalsize 
}

\baselineskip= 2\baselineskip
\FRAME{ftbpFO}{0.0277in}{0.0277in}{0pt}{\Qct{The variation of the energy
spectrum in units of $fm^{-1}$ versus the screening parameter $\protect%
\delta $.\newline
}}{}{Figure 1}{}\FRAME{ftbpFO}{0.0277in}{0.0277in}{0pt}{\Qct{The variation
of the energy spectrum versus the mass $\protect\mu $. All parameters are in
units of $\ fm^{-1}$.\newline
}}{}{Figure 2}{}\FRAME{ftbpFO}{0.0277in}{0.0277in}{0pt}{\Qct{The variation
of the energy spectrum versus the constant $C$. All parameters are in units
of $fm^{-1}$.}}{}{Figure 3}{}

\begin{table}[tbp]
\caption{{\protect\small The negative-energy degenerate states in units of $%
fm^{-1}$ of the pseudospin-symmetry Hulth$\mathbf{{\acute{e}}}$n potential
for various values of $n_{r}$, $\tilde{\ell}$ and $\protect\delta $. For a
special case, $\protect\mu =5~fm^{-1}$, $\Delta _{0}=3.4~fm^{-1}$ and $%
C=-4.9~fm^{-1}$}.\newline
}%
\begin{tabular}{llllllllll}
\tableline\tableline$\tilde{\ell}$~~ & ~$n_{r}$~ & ~$\delta $~~ & 
Degenerate~States & ~$E_{n_{r},\kappa }$ & $\tilde{\ell}$~ & ~$n_{r}$ & ~$%
\delta $ & Degenerate~States & ~$E_{n_{r},\kappa }$ \\ 
\tableline$1$ & $1$ & $0.025$ & $(1s_{1/2},~0d_{3/2})$ & $0.0963638$ & $1$ & 
$2$ & $0.025$ & $(2s_{1/2},~1d_{3/2})$ & $0.0928939$ \\ 
&  & $0.100$ &  & $0.0425738$ &  &  & $0.100$ &  & $-0.0103694$ \\ 
&  & $0.175$ &  & $-0.0710009$ &  &  & $0.175$ &  & $-0.2174930$ \\ 
&  & $0.250$ &  & $-0.2346580$ &  &  & $0.250$ &  & $-0.4920870$ \\ 
$2$ & $1$ & $0.025$ & $(1p_{3/2},~0f_{5/2})$ & $0.0912282$ & $2$ & $2$ & $%
0.025$ & $(2p_{3/2},~1f_{5/2})$ & $0.0863238$ \\ 
&  & $0.100$ &  & $-0.0363590$ &  &  & $0.100$ &  & $-0.1078600$ \\ 
&  & $0.175$ &  & $-0.2930130$ &  &  & $0.175$ &  & $-0.4732160$ \\ 
&  & $0.250$ &  & $-0.6351320$ &  &  & $0.250$ &  & $-0.9131390$ \\ 
$3$ & $1$ & $0.025$ & $(1d_{5/2},~0g_{7/2})$ & $0.0839128$ & $3$ & $2$ & $%
0.025$ & $(2d_{5/2},~1g_{7/2})$ & $0.0775818$ \\ 
&  & $0.100$ &  & $-0.1447100$ &  &  & $0.100$ &  & $-0.2316110$ \\ 
&  & $0.175$ &  & $-0.5760950$ &  &  & $0.175$ &  & $-0.7705370$ \\ 
&  & $0.250$ &  & $-1.0984500$ &  &  & $0.250$ &  & $-1.3540100$ \\ 
$4$ & $1$ & $0.025$ & $(1f_{7/2},~0h_{9/2})$ & $0.0744360$ & $4$ & $2$ & $%
0.025$ & $(2f_{7/2},~1h_{9/2})$ & $0.0666955$ \\ 
&  & $0.100$ &  & $-0.2784550$ &  &  & $0.100$ &  & $-0.3771030$ \\ 
&  & $0.175$ &  & $-0.8953110$ &  &  & $0.175$ &  & $-1.0870200$ \\ 
&  & $0.250$ &  & $-1.5671200$ &  &  & $0.250$ &  & $-1.7758200$ \\ 
\tableline &  &  &  &  &  &  &  &  & 
\end{tabular}%
\end{table}

\bigskip

\end{document}